\documentclass[traditabstract,rnote]{aa}  

\usepackage{amssymb}
\usepackage{graphicx}
\usepackage{amsmath}
\usepackage{natbib}
\usepackage{txfonts}
\usepackage{fixltx2e}
\usepackage{wasysym}
\usepackage{color}

\begin{document}

\def\na{New A, }%
\def\apj{ApJ, }%
\def\aap{A\&A, }%
\def\aapr{A\&ARv, }%
\def\apjs{ApJS, }%
\def\apjl{ApJ, }%
\def\aj{AJ, }%
\def\pasp{PASP, }%
\def\ssr{Space~Sci.~Rev., }%
\def\prd{Phys.~Rev.~D, }%
\def\mnras{MNRAS, }%
\def\memsai{Mem.~Soc.~Astron.~Italiana, }%
\def\araa{ARA\&A, }%
\def\nat{Nature, }

\def\pasj{PASJ, }

\def\gammaray{$\gamma$-ray}
\def\gammarays{$\gamma$-rays}
\def\gaga{$\gamma-\gamma$}
\def\hess{H.E.S.S.}
\def\chandra{\emph{Chandra}}
\def\xmm{\emph{XMM-Newton}}
\def\asca{\emph{ASCA}}
\def\fermi{\emph{Fermi}-LAT}
\def\deg{$^{\circ}$}

\newcommand{\IS}[1]{\textcolor{red}{\textbf{\boldmath#1\unboldmath}}}

\title{Gravitational Light Bending Prevents $\gamma\gamma$ Absorption in Gravitational
Lenses}


\author{Markus B\"ottcher \inst{1,2}
\and
Hannes Thiersen \inst{1}}
\institute{Centre for Space Research, North-West University, Potchefstroom 2520, South Africa
\and
Astrophysical Institute, Department of Physics and Astronomy, Ohio University, Athens, OH 45701, USA
}
\date{Received ...; accepted ...}
\abstract
{The magnification effect due to gravitational lensing enhances the chances of detecting moderate-redshift
($z \sim 1$) sources in very-high-energy (VHE; $E > 100$~GeV) $\gamma$-rays by ground-based Atmospheric 
Cherenkov Telescope facilities. It has been shown in previous work that this prospect is not hampered by
potential $\gamma-\gamma$ absorption effects by the intervening (lensing) galaxy, nor by any individual
star within the intervening galaxy. In this paper, we expand this study to simulate the light-bending
effect of a realistic ensemble of stars. We first demonstrate that, for realistic parameters of the 
galaxy's star field, it is extremely unlikely (probability $\lesssim 10^{-6}$) that the direct line of 
sight between the $\gamma$-ray source and the observer passes by any star in the field close enough to 
be subject to significant $\gamma\gamma$ absorption. Our simulations then focus on the rare cases where 
$\gamma\gamma$ absorption by (at least) one individual star might be non-negligible. We show that 
gravitational light bending will have the effect of avoiding the $\gamma-\gamma$ absorption spheres 
around massive stars in the intervening galaxy. This confirms previous results by 
Barnacka et al. and re-inforces arguments in favour of VHE 
$\gamma$-ray observations of lensed moderate-redshift blazars to extend the redshift range of 
objects detected in VHE $\gamma$-rays, and to probe the location of the $\gamma$-ray emission 
region in those blazars.}


\authorrunning{H. Thiersen \& M. B\"ottcher}
\titlerunning{Gamma-gamma absorption in lensing galaxies}
\maketitle

\section{\label{intro}Introduction}

To date, about 40 blazars (jet-dominated active galactic nuclei with their relativistic jets oriented at
a small angle with respect to the line of sight) have been detected by ground-based Atmospheric Cherenkov
Telescope facilities as sources of very-high-energy (VHE; $E > 100$~GeV) $\gamma$-rays\footnote{see
{\tt http://tevcat.uchicago.edu}}. Their distances span a redshift range $0 < z < 0.944$. This range is 
primarily limited by the $\gamma-\gamma$ absorption effect of the Extragalactic Background Light (EBL) 
on VHE $\gamma$-rays from cosmological distances 
\citep[see, e.g.,][]{stecker92,dejager94,dwek05,franceschini08,finke10,dominguez13}. To
expand the $\gamma$-ray horizon set by $\gamma-\gamma$ absorption, sources at higher redshifts either
need to be unusually bright in VHE $\gamma$-rays (exhibiting an unusually hard $\gamma$-ray spectrum).
Alternatively, VHE blazars of known classes can be gravitationally lensed, whereby their observed
fluxes are magnified. Two $\gamma$-ray blazars detected by the {\it Fermi} Large Area Telescope
({\it Fermi}-LAT) are known to be gravitationally lensed, namely PKS 1830-211 \citep{barnacka11}
and S3~0218+357 \citep{cheung14}. The latter has been successfully detected at VHE $\gamma$-rays
by the Major Atmospheric Cherenkov Telescope \citep[MAGIC, ][]{mirzoyan14}, thus making it the
most distant known VHE $\gamma$-ray emitter to date at $z = 0.944$. 

The $\gamma$-ray detection of gravitationally lensed blazars not only promises the extension of
the VHE blazar catalogue to higher redshifts. \cite{barnacka14a,barnacka15} have shown that the
time delay between the two images of a gravitationally lensed blazar depends very sensitively on
the exact location of the emission region in the source plane. Thus, differences in the locations
of the emission region dominating the variable emission at different frequency ranges (e.g., radio 
vs. optical vs. $\gamma$-rays) may lead to different time delays between the two images in those
different frequency bands. Hence, such differences in time delays may be used to probe the location
of the $\gamma$-ray emission region in blazars, relative to the radio or optical emission region
\citep{barnacka14a,barnacka15}. Note that this method may be applied even when the two lensed
images are not spatially resolved, by searching for repeating variability patterns corresponding
to the two lensed images. 

An important question concerning the feasibility of such studies is whether the additional 
IR -- optical -- UV radiation field provided by the lensing galaxy and its stellar population
may provide a significant source of $\gamma-\gamma$ opacity, thus effectively preventing the
VHE $\gamma$-ray detection of lensed blazars in significant numbers. To investigate this, 
\cite{barnacka14b} have calculated the $\gamma-\gamma$ opacity of the collective radiation
field of a typical $L_{\ast}$ galaxy as well as the opacity provided by an individual star
within the lensing galaxy. In both cases, they found that these intervening sources of soft
radiation field do not lead to significant $\gamma-\gamma$ absorption. Intriguingly, even if
the direct line of sight to the background blazar passes very close to a star in the lensing
galaxy (i.e., closer than the characteristic radius within which the $\gamma-\gamma$ opacity 
exceeds one), the gravitational-lensing effect naturally bends the light path significantly
further away from the star, thus helping to avoid $\gamma-\gamma$ absorption. While this is
an exciting result, suggesting that excess $\gamma-\gamma$ absorption due to the lens is not
a hindrance to VHE detections of distant, gravitationally lensed blazars, their study was
limited to just one individual star in the lensing galaxy; it is not a priori clear whether 
this result still holds if the light-bending effect of a realistic ensemble of (typically 
billions of) stars within the lensing galaxy. 

In this paper, we therefore extend the study of \cite{barnacka14b} to simulate the paths of
$\gamma$-rays through a representative star field in an intervening, lensing galaxy, finding
that even when considering a realistic stellar population in the galaxy, the light bending
effect still aids VHE $\gamma$-rays to avoid the $\gamma-\gamma$ absorption spheres of all
stars in the field. We describe the general model setup and the numerical method to trace
the paths of $\gamma$-rays through the lensing galaxy in Section \ref{setup}. Results are
presented in Section \ref{results}, and we summarize in Section \ref{summary}.

\section{\label{setup}Numerical Setup}

In order to evaluate the effect of gravitational light bending on the path of a VHE $\gamma$-ray
(or any other photon) through an intervening galaxy, we performed ray-tracing simulations. The
general setup of these simulations is as follows: For a generic case, we assume that the lens 
is at a distance of 3~Gpc from the observer on Earth, while the background blazar is located 
a distance of 3~Gpc behind the lens. The average density of stars in the region of the galaxy 
through which the $\gamma$-ray passes, is parameterized as $n_{\ast} = 10^{-2} \, n_{-2}$~ly$^{-3}$.
For $n_{-2} = 1$, this is slightly larger than the density of stars in the solar neighbourhood. 
An approximately uniform distribution is the first, natural, follow-up to the study by
\cite{barnacka14b}, so we will focus on this case here. However, below, we also briefly discuss
a more realistic scenario, in which a much denser star cluster may be located in the $\gamma$-ray
path.

In order to assess whether the $\gamma$-ray passes through any $\gamma-\gamma$ absorption sphere 
(and, thus, may be subject to significant $\gamma-\gamma$ absorption by any of the stellar radiation 
fields), we estimate the radius $r_{\gamma\gamma}$ of the $\gamma-\gamma$ absorption sphere (where 
the $\gamma-\gamma$ opacity $\tau_{\gamma\gamma} = 1$) using $r_{\gamma\gamma} = 10^9 \, (L / L_{\odot}) 
\, E_{\rm eV}^{-1} \; {\rm cm}$ \citep{barnacka14b}, where $L / L_{\odot}$ is the stellar 
luminosity normalized to the solar luminosity, and $E_{\rm eV}$ is the peak photon 
energy of the stellar spectrum. The above expression represents the maximum size of 
the $\gamma-\gamma$ absorption sphere for $\gamma$-ray photons optimally interacting 
with the stellar photons, i.e., $E_{\gamma} = 520 \, E_{\rm eV}^{-1}$~GeV. To express 
$r_{\gamma\gamma}$ solely as a function of stellar mass, $m \equiv M / M_{\odot}$, we 
use a scaling of the stellar luminosity as $L = L_{\odot} \, m^{3.5}$
\citep[which is a convenient interpolation between a slightly shallower mass dependence
at low masses and a slightly steeper one at larger masses, e.g.,][]{DK91} and peak photon 
energy $E_{\rm eV} = 0.5 \, m^{0.5}$, yielding finally

\begin{equation}
r_{\gamma\gamma} = 2 \times 10^9 \, m^3 \; {\rm cm}
\label{rgg2}
\end{equation}

We first estimate the probability of a $\gamma$-ray being subject to significant $\gamma-\gamma$
absorption in the star field of a galaxy, irrespective of any gravitational light bending effects.
We consider $\gamma-\gamma$ absorption to be non-negligible of $\tau_{\gamma\gamma} \ge 0.1$. Due
to the scaling of $\tau_{\gamma\gamma} \propto b^{-1}$, with $b$ being the impact parameter (i.e., 
the distance of closest approach of the direct line of sight of the $\gamma$-ray to the star),
this means, a star of mass $m$ has an effective cross section of $\sigma = 100 \, \pi \, 
r_{\gamma\gamma}^2$. Assuming an approximately constant stellar density over a scale height
$h = 1 \, h_{kpc}$~kpc of the galaxy, the probability of a $\gamma$-ray passing within $10 \,
r_{\gamma\gamma}$ of any star, is then 

\begin{equation}
P_{\gamma\gamma} = 100 \, \pi \, h \int\limits_{m_1}^{m_2} dm \, n(m) \, r_{\gamma\gamma}^2 (m)
\label{P1} 
\end{equation}
where $n(m)$ is the mass distribution of stars. For the purpose of this simple estimate, we 
approximate the stellar mass function as a single power-law $n(m) = n_0 \, m^{-\alpha_{m}}$
with $\alpha_m = 2.5$ between $m_1 = 0.08$ and $m_2 = 100$. This then yields a probability
of $P_{\gamma\gamma} = 3 \times 10^{-7} \, h_{\rm kpc} \, n_{-2}$. Thus, for realistic
parameters of the stellar field and the scale height of the galaxy, even without the effects
of gravitational light bending, it is extremely unlikely that $\gamma\gamma$ absorption in
the radiation field of any individual star will play a significant role. In the following, 
we consider the rare case in which one of the stars in the field is, by chance, located 
close enough to the direct line of sight from the $\gamma$-ray source to the observer to
cause significant $\gamma\gamma$ absorption if gravitational light bending were not taken
into account.

The deflection angle $\alpha_{\rm def}$ resulting from the $\gamma$-ray passing the star at an
impact parameter $b$, is given by

\begin{equation}
\alpha_{\rm def} = {4 \, G \, M \over c^2 \, b}
\label{alpha}
\end{equation}
where $M$ is the mass of the star, and is generally $\alpha_{\rm def} \lesssim 10^{-6}$ 
for main-sequence stars if $b$ is larger than the stellar radius. Consequently, the total 
deflection of the $\gamma$-ray even when interacting with thousands of stars, is very small. 
We therefore restrict our simulations to a cylinder of radius $R = 10$~ly and height $h = 1$~kpc, 
as a characteristic scale height of the galaxy. Thus, the gravitational influence of stars further
than 5~ly away from the direct line of sight is neglected. Within our simulation volume, we 
randomly distribute stars with an average number density $n_{\ast}$, except for placing one 
randomly chosen star deliberately close (at $b < r_{\gamma\gamma}$) to the direct line of 
sight in order to investigate the rare cases where $\gamma\gamma$ absorption would be 
relevant without gravitational light deflection. 

The masses of the stars are 
randomly drawn from a Salpeter mass function, $N_{\ast} (M) \propto M^{-2.5}$. As has been
shown in \cite{barnacka14b}, low-mass stars (less than a few $M_{\odot}$) have too low luminosity
to cause any significant $\gamma-\gamma$ absorption. Furhermore, due to their small masses, any 
angular deflection caused by them is also expected to be negligible. In our simulations, we 
therefore restrict the considered mass range to $1 \, M_{\odot} \le M \le 100 \, M_{\odot}$, 
neglecting the influence of low-mass stars. This reduces the number density of stars actually
considered in the simulations. Specifically, we conservatively include 250 stars in the mass 
range $1 \, M_{\odot} \le M \le 100 \, M_{\odot}$ within the simulation volume.

\begin{figure}
\vskip -3.5cm
\centering
\resizebox{\hsize}{!}{\includegraphics{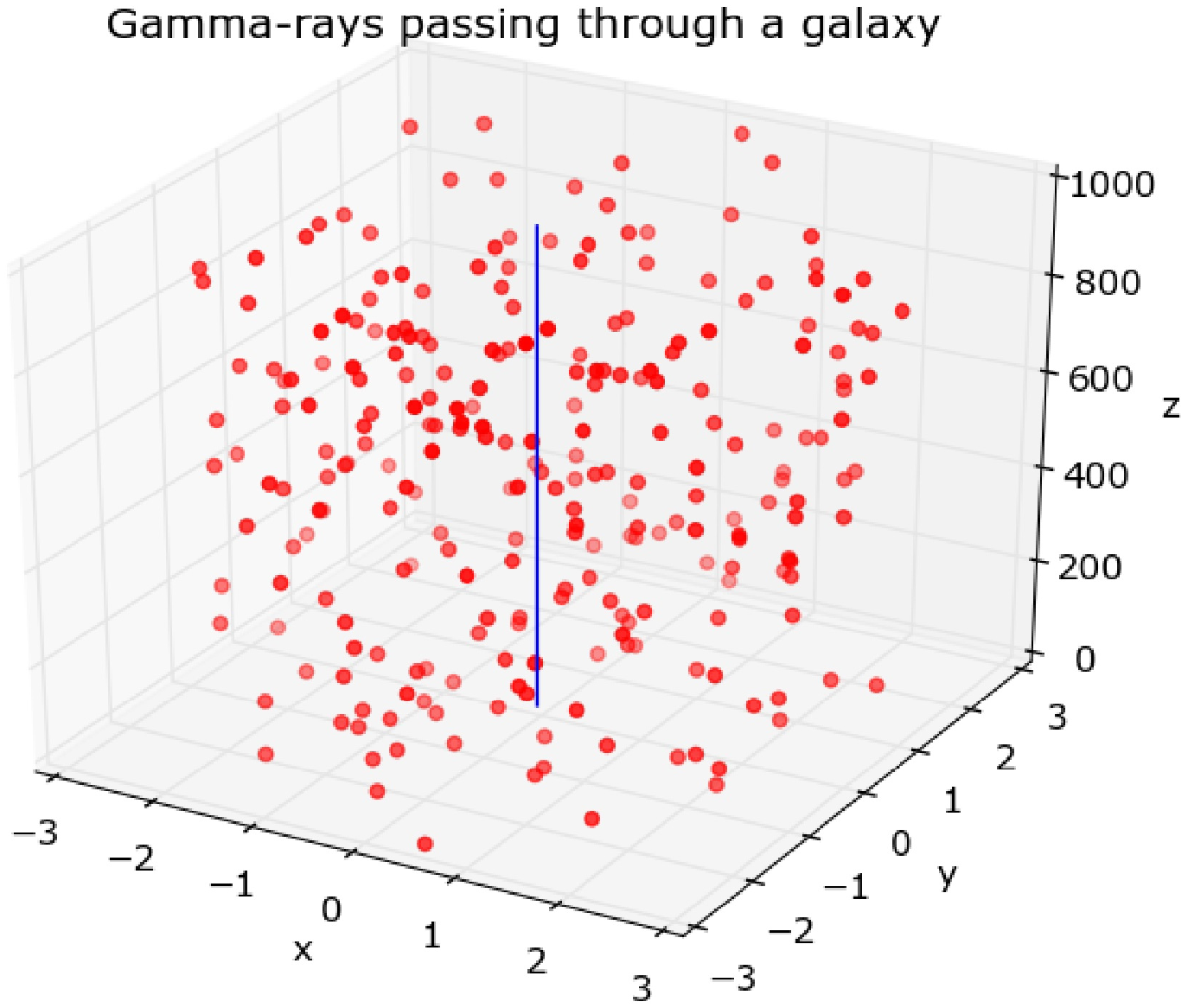}}
\vskip -3.5cm
\caption{Illustration of a typical example of our $\gamma$-ray tracing simulations. Red dots mark 
the positions of stars; the blue line indicates the path of the $\gamma$-ray which continues to 
propagate towards the observer. The axes are labelled in units of pc.}
\label{path}
\end{figure}

Our ray-tracing code then scans through a fine grid of $\gamma$-ray photon arrival directions
into the simulation volume, to find the path which ultimately propagates to the observer on Earth
(while most photon paths are being deflected in other directions which will miss the observer).
Figure \ref{path} illustrates the observed $\gamma$-ray photon path for one of our simulations.
The code tracks the distances of closest approach to each star in the simulation volume. 

For every star $i$ in the simulation volume, the impact parameter $b_i$ is then normalized to 
$r_{\gamma\gamma}^i$ of the star, to check whether the $\gamma$-ray photon path traverses or 
avoids that star's $\gamma-\gamma$ absorption sphere. The simulation is repeated 100 times with
different random seeds (to determine the stars' positions and masses), in order to assess the 
statistical significance of our results.

\begin{figure}
\centering
\resizebox{\hsize}{!}{\includegraphics{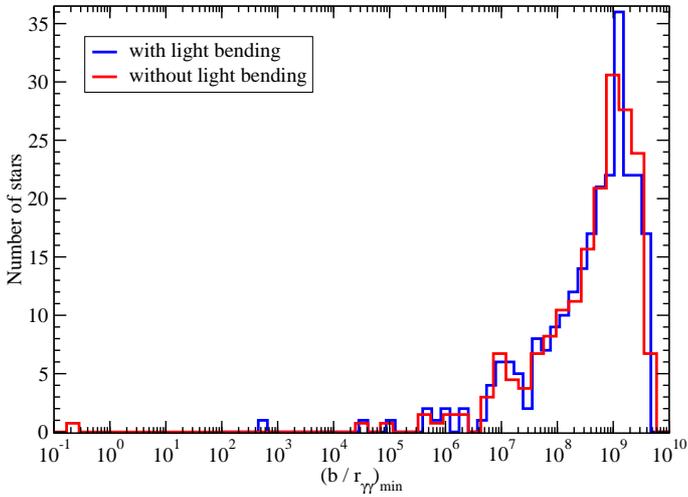}}
\caption{Histogram of normalized impact parameters ($b_i / r_{\gamma\gamma}^i$) for all 250 stars
in one representative simulation. The blue histogram illustrates the result of the ray-tracing,
including gravitational light bending, while the red histogram shows the result corresponding to
a straight photon path without light bending (re-normalized to account for the different bin
sizes of the two histograms). }
\label{impacts}
\end{figure}

\section{\label{results}Results}

Figure \ref{impacts} shows a histogram of the impact parameters (normalized to the $\gamma$-ray
absorption sphere) of the $\gamma$-ray photon reaching the observer on Earth, for all 250 stars 
in one of our simulations (blue histogram). It illustrates that the observed $\gamma$-ray passes 
no star in in the simulation at a distance less than several hundred times $r_{\gamma\gamma}$. 
This is compared to the result corresponding to a straight
photon path along the $z$ axis, i.e., what would be expected without gravitational light-bending
effects. It is clear that the gravitational light bending systematically shifts the impact parameters
to larger values and, in particular, shifts the minimum impact parameter to a value clearly outside
the $\gamma\gamma$ absorption sphere. This suggests that even with a realistic ensemble of stars, 
the gravitational light bending tends to aid $\gamma$-ray photons to avoid the $\gamma\gamma$ 
absorption spheres of stars potentially providing a significant $\gamma\gamma$ opacity. 

\begin{figure}
\centering
\resizebox{\hsize}{!}{\includegraphics{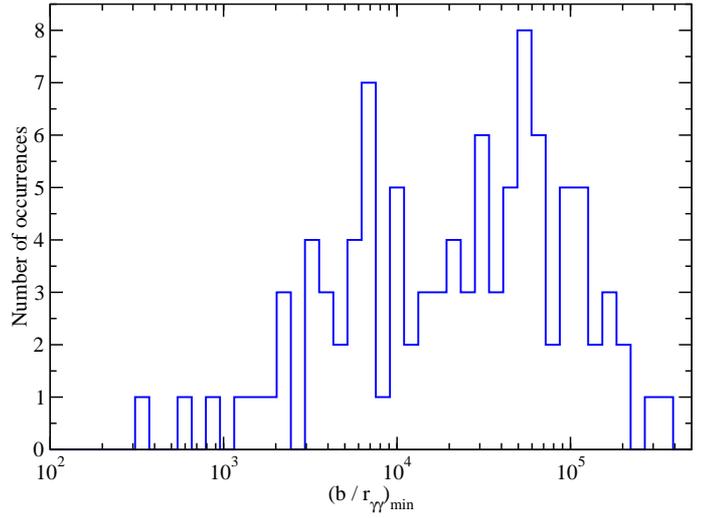}}
\caption{Histogram showing the minimum normalized impact parameters ($b_i / r_{\gamma\gamma}$) 
for all 100 Monte-Carlo realizations of our simulation. }
\label{bmin}
\end{figure}

Clearly the most important result of our simulations is the minimum normalized impact parameter
$b_i / r_{\gamma\gamma}$, i.e., the distance of closest approach, relative to the star's $\gamma-\gamma$
absorption sphere, to {\it any} star in the ensemble. Figure \ref{bmin} shows a histogram of the 
minimum $b_i / r_{\gamma\gamma}$ for each of the 100 Monte-Carlo realizations of our simulations.
It illustrates that in no case will any star be passed at a closer range than a few hundred times 
the $\gamma\gamma$ absorption sphere radius. Consequently, $\gamma$-rays will never be subject to
$\gamma\gamma$ opacities larger than $\sim 10^{-2}$. This confirms our conclusion that, irrespective 
of the details of the stellar distribution in the intervening galaxy, even if the direct line of sight 
were to pass very close (within the $\gamma\gamma$ absorption sphere) of any star in the galaxy, 
the light bending effect will act to help VHE $\gamma$-rays to avoid the $\gamma\gamma$ absorption 
spheres of all stars. 

The uniform star distribution used in the above considerations, is certainly an over-simplification
of the actual distribution of stars in a galaxy. In order to investigate the effects of a more realistic,
inhomogeneous star distribution, we have also explored situations in which (a) a typical O/B association
located within the disk of the galaxy and (b) a globular cluster in the galactic halo are located in the
direct line of sight. In both cases, we find results consistent with those displayed in Figures \ref{impacts}
and \ref{bmin}. For the case of a globular cluster in the line of sight, we note that, while the probability
of a close encounter of the $\gamma$-ray with a star is very high, globular clusters only contain low-mass
stars (typically type G and later), which, due to their low luminosities, never provide a significant 
$\gamma\gamma$ opacity, irrespective of the impact parameter \citep{barnacka14b}.

\section{\label{summary}Summary}

We have re-evaluated the result of \cite{barnacka14b} that VHE $\gamma$-rays from a gravitationally
lensed blazar are not expected to be subject to significant $\gamma\gamma$ absorption by the radiation
field of the lens, because the gravitational light bending effect will cause the $\gamma$-ray paths
to systematically avoid the $\gamma\gamma$ absorption spheres of lensing systems. In \cite{barnacka14b},
only the collective radiation field of an entire galaxy and the radiation fields of one individual star
within a lensing galaxy were considered. As this might be over-simplifying the situation present for
$\gamma$-rays passing through the potentially dense star field of a galaxy, we have evaluated the
light bending effect due to a realistic stellar population in a lensing galaxy. 

We have first evaluated the probability of {\it any} star to be close enough to the direct line of
sight between a background VHE $\gamma$-ray source and the observer to cause significant $\gamma\gamma$
absorption if light bending were not taken into account. We find this probability to be very low,
typically $\lesssim 10^{-6}$. We then concentrated on the few exceptional cases in which a star
might be, by chance, located very close to the line of sight. For those cases, we have shown that
the result of \cite{barnacka14b} still holds, namely that the gravitational light bending effect
will deflect the $\gamma$-ray path far beyond the $\gamma\gamma$ absorption sphere of that single
star, without causing it to approach any other star close to its $\gamma\gamma$ absorption sphere. 

These results confirm the findings of \cite{barnacka14b} and
reinforce the prospect to detect gravitationally lensed $\gamma$-ray blazars with
ground-based VHE $\gamma$-ray observatories, especially the future Cherenkov Telescope Array (CTA).
As suggested by \cite{barnacka14a,barnacka15}, the measurement of time delays between lensed 
images (which can not be directly spatially resolved at $\gamma$-ray energies) will then allow
one to probe the location of the $\gamma$-ray emission region in comparison to the lower-energy
emission.

\acknowledgements{ We thank the anonymous referee for a constructive and helpful report.
MB acknowledges support the South African Research Chair Initiative (SARChI) of the
Department of Science and Technology and the National Research Foundation\footnote{Any opinion, finding 
and conclusion or recommendation expressed in this material is that of the authors and the NRF does not 
accept any liability in this regard.} of South Africa, under SARChI Chair grant No. 64789.}







\end{document}